\documentclass[10pt,aps,pre,noshowpacs,twocolumn,superscriptaddress,nobibnotes]{revtex4-1}
\usepackage{amssymb}
\usepackage{graphicx}
\usepackage{amsmath,amsfonts}
\usepackage[colorlinks=true,allcolors=blue]{hyperref}
\usepackage[utf8]{inputenc}
\usepackage{xcolor}  
\usepackage[normalem]{ulem}
\usepackage[T1]{fontenc} 
\usepackage{multirow}
\usepackage{booktabs}
\usepackage{threeparttable}
\usepackage{tabularx}
\usepackage{adjustbox}
\usepackage{float}

\usepackage{pdfpages}
\makeatletter
\AtBeginDocument{\let\LS@rot\@undefined}
\makeatother

\begin{document}

	\title{Feature-aware ultra-low dimensional reduction of real networks}
	
	\date{\today}
	
	\author{Robert Jankowski}
	\affiliation{Departament de F\'isica de la Mat\`eria Condensada, Universitat de Barcelona, Mart\'i i Franqu\`es 1, E-08028 Barcelona, Spain}
	\affiliation{Universitat de Barcelona Institute of Complex Systems (UBICS), Universitat de Barcelona, Barcelona, Spain}
	\author{Pegah Hozhabrierdi}
	\affiliation{Department of Electrical Engineering and Computer Science, Syracuse University, 223 Link Hall, Syracuse, NY 13244}%
	\author{Mari\'an Bogu\~{n}\'a}
	\affiliation{Departament de F\'isica de la Mat\`eria Condensada, Universitat de Barcelona, Mart\'i i Franqu\`es 1, E-08028 Barcelona, Spain}
	\affiliation{Universitat de Barcelona Institute of Complex Systems (UBICS), Universitat de Barcelona, Barcelona, Spain}
	\author{M. {\'A}ngeles Serrano}
	\email[]{marian.serrano@ub.edu}
	\affiliation{Departament de F\'isica de la Mat\`eria Condensada, Universitat de Barcelona, Mart\'i i Franqu\`es 1, E-08028 Barcelona, Spain}
	\affiliation{Universitat de Barcelona Institute of Complex Systems (UBICS), Universitat de Barcelona, Barcelona, Spain}
	\affiliation{ICREA, Passeig Llu\'is Companys 23, E-08010 Barcelona, Spain}

	\begin{abstract}
		In existing models and embedding methods of networked systems, node features describing their qualities are usually overlooked in favor of focusing solely on node connectivity. This study introduces $FiD$-Mercator, a model-based ultra-low dimensional reduction technique that integrates node features with network structure to create $D$-dimensional maps of complex networks in a hyperbolic space. This embedding method efficiently uses features as an initial condition, guiding the search of nodes' coordinates towards an optimal solution. The research reveals that downstream task performance improves with the correlation between network connectivity and features, emphasizing the importance of such correlation for enhancing the description and predictability of real networks. Simultaneously, hyperbolic embedding's ability to reproduce local network properties remains unaffected by the inclusion of features. The findings highlight the necessity for developing network embedding techniques capable of exploiting such correlations to optimize both network structure and feature association jointly in the future.
	\end{abstract}

	\maketitle
	\let\oldaddcontentsline\addcontentsline
	\renewcommand{\addcontentsline}[3]{}

	\section{Introduction}
	Factors relevant to the links' formation process in real-world complex networks are typically numerous, and unknown or difficult to measure. For instance, off-line and on-line social relations of friendship can be sustained by a variety of reasons including individual factors, such as approachability or social skills, and environmental factors, such as geographic proximity and life events and activities. This has not prevented that generative models based on simple connectivity rules are able to capture many of the salient properties of real networks, such as degree heterogeneity, small-worldness, clustering, and community structure. 
	
	Such is the case of a family of geometric random graph models~\cite{serrano2008similarity,Boguna2021}, in which pairs of nodes are connected with a probability that depends on their popularity and on their distance in a similarity metric space. For a real network, these variables can be inferred by maximizing the likelihood of the model to reproduce the observed structure, as done by the embedding tool Mercator~\cite{Garcia2019}. The combined use of geometric models and embedding tools prove that hyperbolic space is a natural geometry for real networks~\cite{krioukov2010hyperbolic}, and have been able to explain heterogeneous degree distributions, clustering, small-worldness, percolation, spectral properties, community structure, different forms of self-similarity, preferential attachment in growing networks, the non-trivial coupling between weights and topology, correlations in multilayer networks, and served as the basis for defining a renormalization group for complex networks, see \cite{boguna2021network} and references therein.

	The popularity component of the hyperbolic space where real networks are embedded accounts for nodes' degrees, and distances in the similarity subspace for affinities not influenced by degrees. Assuming that relational properties of nodes, other than popularity, define this similarity subspace, it is expected that the organization of real networks often correlates with node descriptive metadata. This acts as additional information to the network topology, enriching the description of nodes and helping to discover, identify, and categorize them. Going back to the example of friendship networks, link formation could be strongly affected by age, gender, interests, background, and other social determinants~\cite{hays1985longitudinal, HAYNIE2014, Bigelow1977}. This influence is often captured in hyperbolic maps of real networks, even though the corresponding information was not deliberately incorporated during the mapping process. For instance, autonomous systems form clusters in the similarity space according to geographic data in hyperbolic maps of the Internet~\cite{Boguna2010}, and functional modules form clusters in the hyperbolic maps of brain connectomes~\cite{allard2020navigable}. However, this might not always be the case and descriptive metadata could be irrelevant or only partially correlated with network structure and cannot be treated as ground truth~\cite{Peel2017}. 
	
	Under certain conditions, integrating metadata with network structure can enhance the accuracy of community detection methods~\cite{Newman2016, Peel2017, Bassolas2022, Mucha2019, Smith2017} and facilitate a more realistic assessment of the robustness of interconnected systems~\cite{Artime2021}. This poses the question of whether integrating structural information and metadata in network models enables a better understanding of connectivity in complex systems and a better performance in downstream tasks like node classification or prediction of missing links. From a broader perspective, adding features to the network structure leads to a hybridization that reveals a distinct type of spatial description through the implementation of a different inductive bias. This bias involves a set of assumptions that consider both features and network structure simultaneously and independently of whether they are correlated.
	
	Embedding techniques in machine learning, particularly graph neural networks (GNNs)~\cite{kipf2017semisupervised,velivckovic2017graph,hamilton2017inductive,xu2018powerful}, naturally work by aggregating the node-level information extracted from the nodes' features via message passing. They have been shown to achieve good performance on downstream tasks. However, vanilla GNNs work well only if a network is homophilic, i.e., a network where connected nodes have the same class labels and similar features~\cite{ma2021homophily,zhu2020beyond}. Many popular GNNs fail to generalize to this setting and are even outperformed by models that ignore the graph structure. It is also worth mentioning that GNNs are data-driven and task-oriented embedding methods. On the other hand, unsupervised machine learning models, such as Node2Vec~\cite{grover2016node2vec} or DeepWalk~\cite{perozzi2014deepwalk}, can be applied to multiple downstream tasks but rely purely on the structure of the graph to generate embeddings.
	
	The integration of network structure and descriptive metadata can be implemented in the embedding process to produce maps of real networks where coordinates of nodes are determined by the two sources of information, and where the relative attention paid to one versus the other can be controlled. The last and more advanced generator of model-based hyperbolic maps is $D$-Mercator~\cite{jankowski2023dmercator}, a generalization of Mercator~\cite{Garcia2019} to multidimensional hyperbolic embedding. As demonstrated in~\cite{Almagro:2021}, real-world complex networks exhibit a compelling fit within the framework of hyperbolic spaces characterized by low dimensionality, typically falling within the range of $D \in [2, 10]$. This observation provides a solid rationale for extending the foundation laid by $D$-Mercator as outlined in~\cite{jankowski2023dmercator}, by incorporating node features into the model.
	
	In this paper, we introduce a model-based dimensional reduction technique named Feature-initialized D-dimensional Mercator, $FiD$-Mercator, that embeds networks into an ultra-low dimensional hyperbolic space by combining network topology and the collection of features describing the qualities of nodes. Our overarching goals are to provide an embedding tool that hybridizes network topology and features, and work towards clarifying to what extent node features can help in explaining the structure of real networks, and under which conditions it is advantageous to supplement geometric network models with this information for specific tasks. We used our tool to analyze link prediction and node classification in real networks and we found that the amount of correlation between network connectivity and descriptive metadata determines performance. In general, when this correlation is high, adding features opens a fast track to optimized embeddings for link prediction and improves the node classification results significantly.
	
	\section{Results}
	\subsection{$FiD$-Mercator embedding method}
	Given a graph-structured dataset consisting of a set of $N$ nodes forming a complex network and a set of $N_F$ features associated with the same set of nodes, the embedding method $FiD$-Mercator finds the hidden variables in the geometric soft configuration $\mathbb{S}^D$ model~\cite{Serrano2008}, $\mathbb{H}^{D+1}$ in the isomorphic hyperbolic geometry formulation~\cite{budel2020random}, that best reproduces the topology of the network. In the $\mathbb{S}^D$ model, networks exist in an underlying metric space with an effective hyperbolic geometry~\cite{Krioukov2009}. 
	
	The coordinates of the $N$ nodes in the latent space represent popularity and similarity dimensions. The popularity dimension of a node is associated with its hidden degree $\kappa$, equivalent to a radial coordinate in hyperbolic space and proportional to its observed degree in the network topology. The similarity subspace is represented as a $D$-sphere, which is a hypersphere in the $(D+1)$-dimensional Euclidean space. In this representation, nodes occupy predetermined positions, with angular distances between them representing all the influential factors, aside from degrees, that affect their likelihood to form connections. 
	
	Connection probabilities are dictated by effective distances, which are derived from both hidden degrees and angular separations. This follows a gravity law principle, whereby nodes with higher hidden degrees or those positioned closely in the similarity space are more likely to connect with others. The connection probability is influenced by two parameters: $\beta$, which must be greater than $D$, controls the clustering of the network ensemble and quantifies the level of coupling between the network topology and the metric space; and $\mu$, which dictates the average degree. In this work, we restrict to $D=2$ for ease of visualization and to constrain the number of variables involved. The similarity space of the $\mathbb{S}^2$ model is represented as a two-dimensional sphere of radius $R$, see Section~\ref{sec:model} for technical details. It is important to note that not all networks are optimally represented in $D=2$. However, the results could only be enhanced by identifying the most suitable dimension for a given network.
	
	On top of leveraging the capacity of the $\mathbb{S}^D$ model to fit realistic network structures, \mbox{$FiD$-Mercator} employs the Uniform Manifold Approximation and Projection (UMAP) algorithm \cite{umap2018} for dimensionality reduction (Section~\ref{sec:umap}). UMAP allows us to obtain an informed initial guess of the coordinates of the nodes in the similarity subspace using the information encoded into the nodes' features. These coordinates are subsequently adjusted by a maximum likelihood estimation technique to improve the probability that the observed network topology is reproduced by the model. Standard $D$-Mercator is instead initialized with coordinates on top of $\mathbb{S}^D$ that only depend on the connectivity structure of the network. Given the complexity of the optimization process, such initial conditions may lead to reasonable but not optimal embeddings. Thus, the key contribution of $FiD$-Mercator over $D$-Mercator is that we determine a guided initial condition by mapping the high-dimensional vectors of node features on the two-sphere instead of extracting it from the network topology, such that the final embedding incorporates information about both the connectivity structure of the network and features describing nodes' qualities. Notice that other manifold learning techniques could be used instead of UMAP as far as they produce a projection in the two-sphere that is able to discern features. UMAP is particularly convenient for its favorable trade-off between accuracy and computation cost.
	
	\textbf{$FiD$-Mercator pipeline.} In practical terms, $FiD$-Mercator builds upon the multidimensional hyperbolic embedding tool $D$-Mercator \cite{jankowski2023dmercator}. This tool employs a machine learning method and a statistical inference technique to pinpoint the hidden degrees and the angular coordinates that maximize the likelihood of accurately reproducing a network's topology through the geometric soft configuration $\mathbb{S}^D$ model. Furthermore, it adjusts the parameters $\beta$ and $\mu$ based on this model. $FiD$-Mercator leverages the maximum likelihood estimation machinery of $D$-Mercator. However, it differs in its approach to obtaining an initial condition from the network topology. Rather than employing the Laplacian Eigenmaps method~\cite{belkin2002laplacian} adjusted by the model, $FiD$-Mercator uses the UMAP dimension reduction technique to embed node features onto the two-sphere. UMAP is a well-established tool for representing and visualizing data in ultra-low dimensions. Offering exceptional efficiency with intuitive parameters, it makes a suitable choice for our applications.
	
	\begin{figure}[t!]
		\centering
		\includegraphics[width=0.45\textwidth]{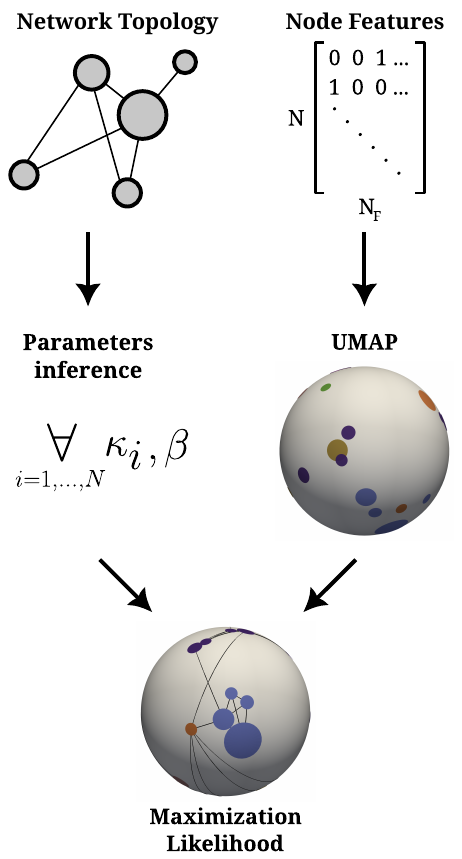}
		\caption{Schematic view on the proposed method \mbox{$FiD$-Mercator}. First, from the network we infer the hidden degrees $\kappa$ and parameter $\beta$. Second, we use the UMAP algorithm to map the nodes onto the two-sphere using the feature matrix. Lastly, the initial nodes' coordinates from the last step are used in the maximization likelihood procedure which tries to fit data to the $\mathbb{S}^2$ model. The size of the nodes is proportional to its expected degree and they are colored according to their communities. Black lines on the two-sphere represent connections produced according to the model.}
		\label{fig:panel0}
	\end{figure}
	
	The $FiD$-Mercator method requires two primary inputs. The first is the adjacency matrix, denoted as $\{a_{ij}\}$, which encodes the topology of the network, with $a_{ij}=1$ if the link between nodes $i$ and $j$ exists and $a_{ij}=0$ otherwise. The second input is the feature matrix, $\{f_{im}\}$. This matrix is formed by the collection of $N_F$ binary features associated with the nodes, where $f_{im}=1$ if node $i$ exhibits feature $m$ and $f_{im}=0$ otherwise. The embedding process consists of the following steps: (i) inferring the hidden degrees $\kappa$ , (ii) inferring the inverse temperature $\beta$ from the network topology, (iii) initial embedding of the nodes using node vector features and the UMAP algorithm, (iv) adjustment of the angular coordinates by maximizing likelihood (ML) to improve the congruency between the original network and the model, and (v) final adjustment of hidden degrees according to the final angular positions. These steps are explained in more detail in Section~\ref{sec:mm}. The schematic representation of the algorithm is shown in Figure~\ref{fig:panel0}.
	
	Notice that by using UMAP, we can work with networks containing more than one connected component. This is a situation that cannot be addressed in $D$-Mercator, as the Laplacian Eigenmaps technique used to provide the initial condition there requires having only one connected component. Nevertheless, in this work we keep working with the largest connected component so that results can be compared with the baseline provided by $D$-Mercator. 
	
	An implementation of $FiD$-Mercator is publicly available at \url{https://github.com/networkgeometry/FiD-Mercator}.
	
	\textbf{Embedding of real networks.}
	
	\begin{figure*}[t]
		\centering
		\includegraphics[width=0.7\textwidth]{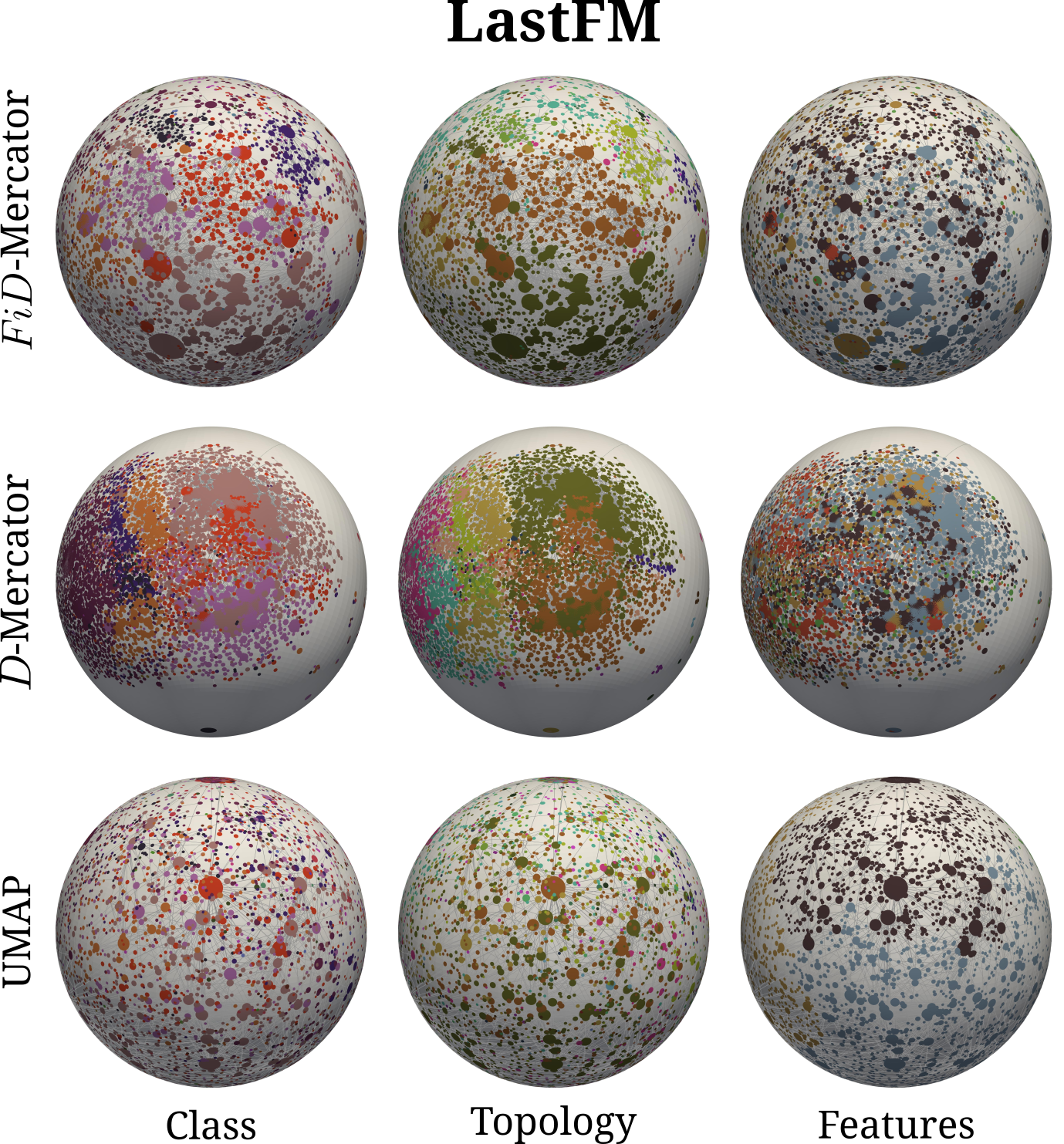}
		\caption{{\bf Two dimensional maps of the LastFM network.} Each row corresponds to the different embedding methods whereas each column is for a different assignment of labels. The size of a node is proportional to its expected degree, and its color indicates the community it belongs to. For the sake of clarity, only the connections with probability (Eq.~\ref{eq:prob_conn}) $p_{ij} > 0.999$ are shown.}
		\label{fig:panel1}
	\end{figure*}

	We gathered ten datasets, which describe structural connectivity and node features of real networks in various domains; these range from citation and social networks to goods and webpage networks. It is noteworthy that a significant portion of the networks we analyzed are commonly used as benchmarks in machine learning research~\cite{kipf2017semisupervised, Pei2020, velickovic2018graph}. Each network is accompanied by descriptive metadata describing the nodes' features as the bag-of-words. A non-zero entry indicates a presence of a given word for the node. For the citation networks, such as Citeseer or Cora, the node represents a publication, and a feature vector is the bag-of-words from the abstract of that publication. On the other hand, for Amazon Photo, the feature vectors are extracted from the product reviews. For more details about the datasets please refer to Section~\ref{sec:datasets}. In Table~\ref{tab:1}, we report global statistics for each network, including number of nodes and features, average degree, average number of features per node, and mean clustering coefficient. 
	
	\begin{figure}[t!]
			\centering
			\includegraphics[width=0.45\textwidth]{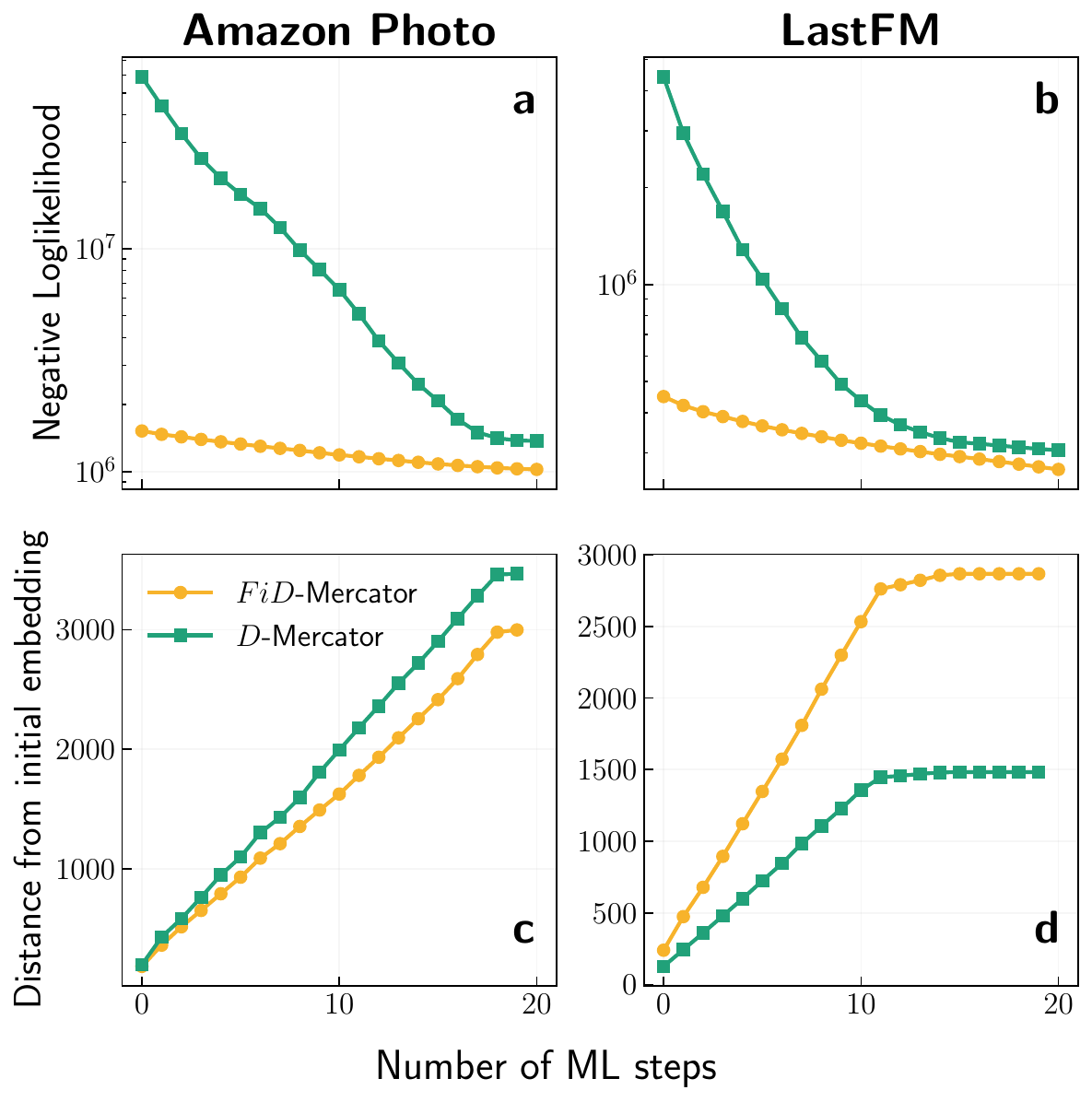}
			\caption{(\textbf{a,b}) Evolution of the global log-likelihood during the maximum likelihood (ML) optimization steps. Each ML step takes a subset of nodes for which new coordinates are proposed. The nodes are ordered through the onion decomposition~\cite{hebert2016multi}. Notice that we plot the negative log-likelihood here, hence the lower the value the better. \textbf{(c,d)} The average angular distance between each node of the initial embedding (for $D$-Mercator: Laplacian Eigenmaps; for $FiD$-Mercator: UMAP) and the embedding after each ML step. We computed the average angular distance only for nodes with $k>2$.}
			\label{fig:panel_loglikelihood}
	\end{figure}

	We produced three different maps for each network: a $FiD$-Mercator embedding that incorporates information about the network topology and the node features; a $D$-Mercator embedding that only relies on the network connectivity structure; and the UMAP projection based solely on the node vector features. As an example, Fig.~\ref{fig:panel1} shows the three embeddings of the LastFM network. Visualizations for the rest of the datasets are given in Supplementary Figs.~12-20. 
	
	The noticeable difference of nodes' layout between the $FiD$-Mercator and $D$-Mercator embeddings, as shown in Fig.~\ref{fig:panel1}, can be quantified by tracking how the global log-likelihood of the embeddings changes during the maximum likelihood optimization steps. Figure~\ref{fig:panel_loglikelihood}(a,b) (see also Supplementary Figs.~1-2) illustrates the time evolution of the global log-likelihood at each step of the  maximum likelihood algorithm. In general, $FiD$-Mercator achieves a higher final value of global log-likelihood compared to $D$-Mercator. Thus, initialization with UMAP can lead to a local optimum that could improve the results of initializing with LE.
	
	Moreover, we carried out experiments to observe how the nodes' coordinates change over time. We computed the average of the angular distances separating the positions of the nodes at $t=0$ and every ML step. Figure~\ref{fig:panel_loglikelihood}(c,d) shows the results for Amazon Photo and LastFM datasets (see also Supplementary Figs.~3-4), where the average is computed for nodes with $k>2$. The figures show that the average angular distance is increasing over time. However, at some moment, it saturates, meaning that the coordinates of nodes with $k>2$ remain stable. This analysis can be used to assess how far the embedding is from the initial condition and specify when to stop the maximum likelihood process.
	
\begin{figure}[t]
		\includegraphics[width=0.5\textwidth]{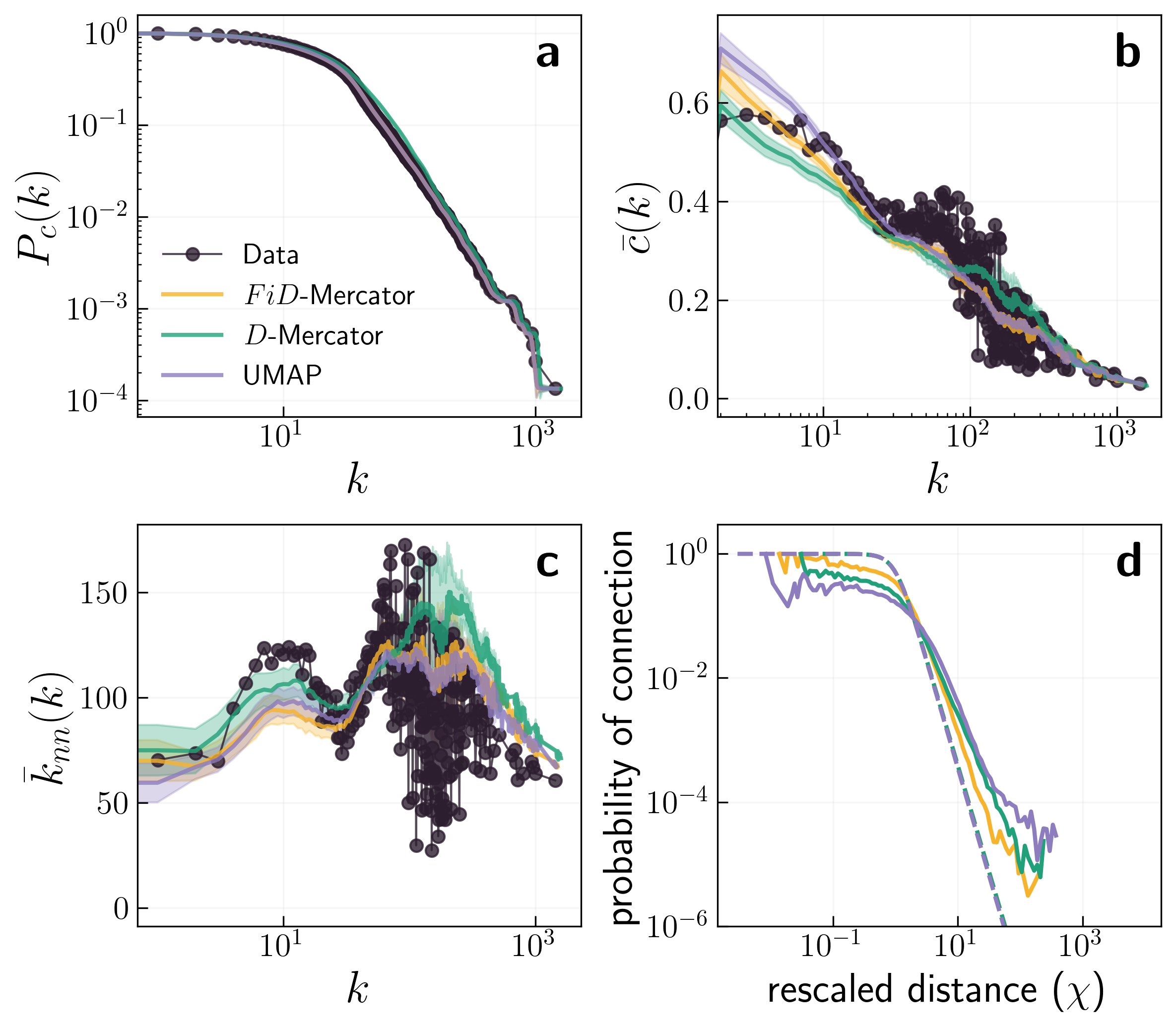}
		\caption{Validation of the embeddings of the Amazon Photo network. Panel (\textbf{a}) shows the complementary cumulative degree distribution and panel (\textbf{b}) the clustering spectrum $\bar{c}(k)$. Symbols correspond to the values in the original network. The lines indicate an estimate of the expected values in the ensemble of random networks in the different embeddings. The $\mathbb{S}^2$ model was used to generate 100 synthetic networks with the parameters and positions inferred by $FiD$-Mercator, $D$-Mercator, or UMAP. The error bars show the $2\sigma$ confidence interval around the expected value. Panel (\textbf{c}) shows the average nearest neighbors degree $\bar{k}_{nn}(k)$. Panel (\textbf{d}) displays the comparison of the expected connection probability based on the estimated $\beta$ (expected) and the actual connection probability computed with the inferred hidden variables.}
		\label{fig:topological_amazon_photo}
	\end{figure}

	The embeddings can be combined with the probability of connection in the $\mathbb{S}^2$ model to generate synthetic surrogates to be compared with the original networks, see Fig~\ref{fig:topological_amazon_photo} and Supplementary Figs.~21-29. One can observe that both $FiD$-Mercator and $D$-Mercator are able to reproduce the degree distribution, the clustering spectrum, and the average nearest neighbours degree with high fidelity. However, these embeddings are not identical and nodes in the $FiD$-Mercator maps show a greater tendency to expand and fill in the similarity subspace. In contrast, the UMAP embedding performs poorly in reproducing network properties as expected, since it contains no information about the network structure. We conclude that adding node features to network structure did not disrupt the quality of the model-based hyperbolic embedding in terms of its capacity at explaining local network topological properties. Next, we are going to explore how adding features can affect downstream tasks that also depend on the mesoscopic and global organization of the networks.
	
	\subsection{Downstream tasks}
	In this section, we show that enriching network topology with node features can have a mixed effect depending on how the two are correlated and on the extent to which each category is relevant for the downstream task to be performed. Next, we report results for link prediction and node classification tasks.
	
	\textbf{Link prediction.}	Given a partially observed network, link prediction (LP) aims to infer the most likely missing links based on the available ones~\cite{LU20111150}. Typically, absent links are ranked in decreasing order of some likelihood value and those at the top are selected as candidates for missing links. The likelihood can be estimated based on some heuristic metric given the patterns of connectivity observed in the network structure or using an underlying network model. When it comes to graph models that employ latent space, the task of link prediction essentially involves ranking pairs of nodes in decreasing order of their connection probability computed using their latent distances. Results from~\cite{Kitsak2020} indicate that as the complexity of a specific link prediction task increases, the utilization of hyperbolic geometry should be more seriously considered.
	
	We conducted tests on the link prediction task in two scenarios. In the first scheme, we randomly removed $L = q \cdot E$ links from the original graph with $q$ defined as the fraction of missing links. Then, we used the coordinates and parameters obtained from the three different embeddings of the complete network to compute the connection probability as given by $\mathbb{S}^2$ (Eq.~\ref{eq:prob_conn}). In the second scheme, we selected the giant connected component (GCC) of the network after removing $L$ links, counted the number of these missing links $L'$ within the GCC, and embedded this subgraph to compute the coordinates of its nodes. We then calculated the likelihood of each non-existing link in the GCC. 
	
	In both cases, we sorted the links in decreasing order of probability, and selected the top $L$ and $L'$, respectively, as predicted links. Even though the Area Under the ROC Curve (AUC) is a popular measure in the literature to assess the goodness of a link prediction method, the precision is a more stringent measure since it concentrates on the top part of the ranking where, under limited resources, more detection efforts would be directed \cite{Linyuan2015,Sun2020,garcia2020precision}. The precision was calculated as the fraction of correctly identified links as presented in~\cite{garcia2020precision}. 
	
	\begin{figure}[t!]
		\centering
		\includegraphics[width=0.5\textwidth]{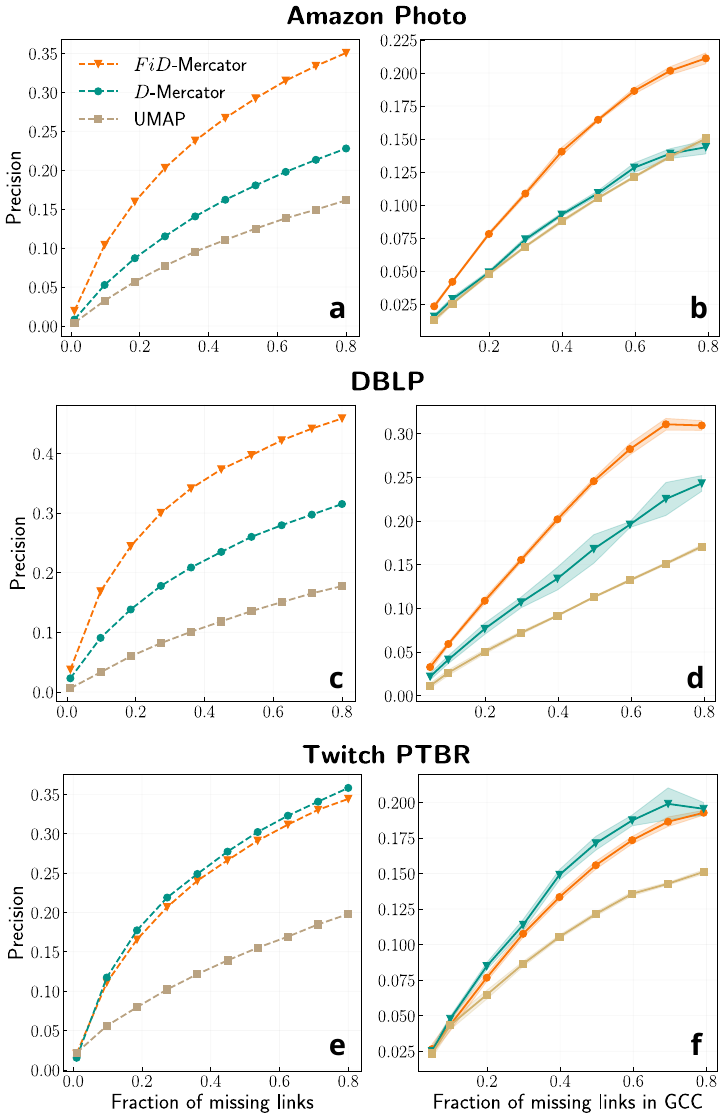}
		\caption{Precision as a function of the fraction of missing links in a link prediction task for (\textbf{a,b}) Amazon Photo, (\textbf{c,d}) DBLP, and (\textbf{e,f}) Twitch PTBR networks. Panels (\textbf{a,c,e}) show the results for the scheme in which the embeddings are of the complete network, whereas panels (\textbf{b,d,f}) are for the second scheme embedding the GCC of the incomplete network. Results are averaged over $5$ different realizations.}
		\label{fig:panel2}
	\end{figure}
	
	The results are reported in Fig.~\ref{fig:panel2} and Supplementary Figs.~9-10. Again, UMAP alone performs poorly whereas $FiD$-Mercator and $D$-Mercator give improved results. In the first scheme, in seven out of the 10 analyzed real networks $FiD$-Mercator outperforms $D$-Mercator by a significant amount. In the remaining three networks, the two schemes reach similar levels of precision. In the second scheme, the overall precision is lower due to the increased difficulty of the task. We excluded three networks (Cornell, Wisconsin, and Texas) from this analysis due to their small size after the removal of links. In three out of the remaining seven networks, $FiD$-Mercator surpasses $D$-Mercator, and in two networks, both methods yield similar results.

	The improvement of $FiD$-Mercator over $D$-Mercator in the LP task can only be explained if node features are correlated with the network topology. To quantify such correlation, we computed the relative difference between the average cosine similarities of features of connected nodes in the original network and features of nodes that were not directly connected (see Section~\ref{sec:correlation} and Table~\ref{tab:1}). As one can notice, higher correlation values for Amazon Photo and DBLP lead to improvement in the link prediction task. On the other hand, when the correlation is weak, both $FiD$-Mercator and $D$-Mercator perform similarly, as in the case of the Twitch PTBR dataset.
	
	\textbf{Node classification.}	Node classification (NC) is a task where the goal is to categorize nodes into different sets described by specific labels. The categories can be based on different sources of information. For each real network, we considered three different ways of assigning labels to nodes. 
	\begin{itemize}
		\item
		{\bf Class.} Using descriptive metadata, different from features, which characterizes some aspect that divides nodes in a limited number of classes, for instance product types, professional roles, research fields, or film genres. The metadata defining the classes is typically provided along the network graph. In this partition, we label nodes with the class names. 
		\item
		{\bf Topology.} By leveraging the network topology to construct the communities of densely connected neighbours. Labels based on topological communities were produced using the Louvain method~\cite{Blondel_2008}, which is a greedy modularity optimization technique. The method is fast and detects the number of communities without predefining it. 
		\item
		{\bf Features.} Based on similarities between node features. Feature-based labels were defined from the UMAP embedding used in $FiD$-Mercator to fix the initial conditions. We applied the agglomerative clustering algorithm from the scikit-learn Python library \cite{scikit-learn} to the UMAP projection. This algorithm is a bottom up hierarchical clustering method in which each node starts in its own cluster, and clusters are successively merged together following a linkage criterium that takes the mean distance between elements of each cluster, also called average linkage clustering used in UPGMA \cite{Sokal1958ASM}. The nodes projected by UMAP in close positions of the similarity space, i.e., separated by a small angular distance, have a tendency to merge into one cluster. To fix the number of labels in the hierarchical clustering algorithm, we employed the geometric community concentration measure used in \cite{jankowski2023dmercator}. This measure computes the level of clusterization of nodes in the underlying two-sphere embedding. The number of clusters is tuned to the one that produces a maximum value of the geometric concentration measure, see Section~\ref{sec:concentration} for more details.
	\end{itemize}

		\begin{figure}[t!]
			\centering
			\includegraphics[width=0.5\textwidth]{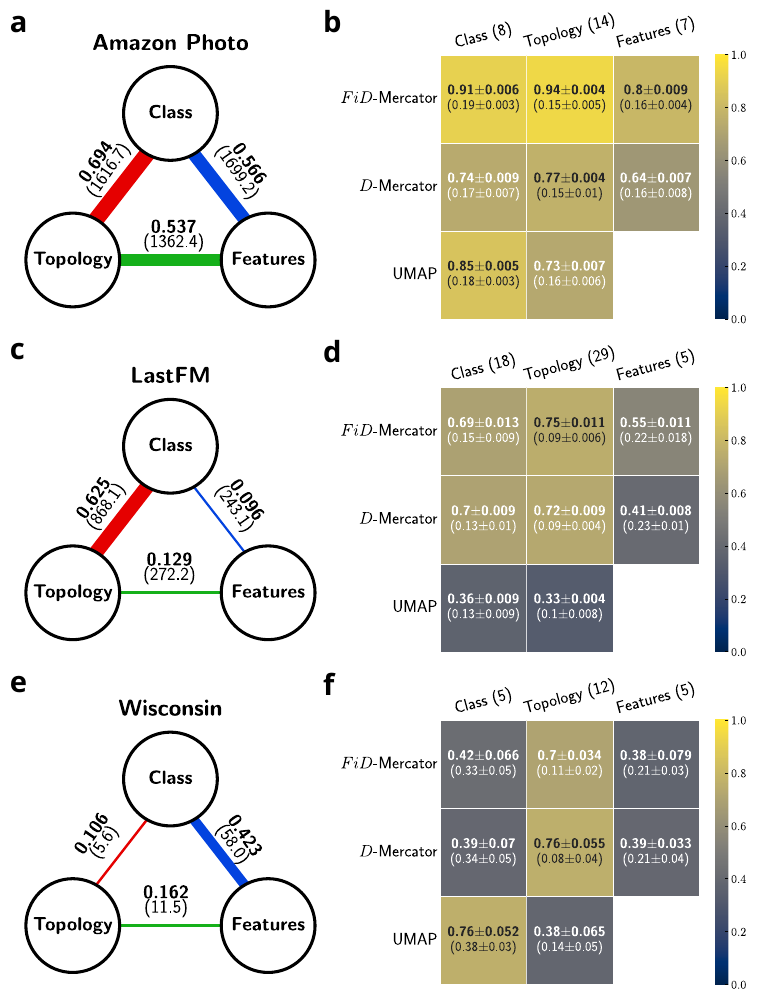}
			\caption{Correlation between node labeling and accuracy in node classification task. (\textbf{a}, \textbf{c}, \textbf{e}) Normalized mutual information (NMI) between each of two different sets of labels. The value in bracket represents the z-score which quantifies the significance of the correlation between two label sets. A higher z-score indicates a stronger and more statistically significant correlation between the two sets of labels (see Section~\ref{sec:random_labels}). The width of the line is proportional to NMI. (\textbf{b}, \textbf{d}, \textbf{f}) Accuracy heatmaps of node classification task. Each row represents a different embedding method, whereas each column a different set of labels. In brackets, the number of labels is shown. The performance of node classification on shuffled labels are displayed in brackets as the second row in each entry. Train-test size split: $80/20$. Results are averaged over $5$ different train-test splits.}
			\label{fig:panel3}
		\end{figure}

		Each node in the real networks was, thus, assigned three labels---a class label, a topological community label, and a feature-based label---, one from each of the three partitionings. This labeling is visualized through color-coding in Figure~\ref{fig:panel1}, which depicts three different maps of the LastFM network, and in Supplementary Figs.~12-20 for the remaining networks. It is worth noting that labels denoting classes and labels based on network topology and on node features do not necessarily correlate with one another. For instance, the absence of correlation is the reason why the positions of nodes in different feature-labeled sets overlap in the embeddings obtained by $FiD$-Mercator and $D$-Mercator.
		
		In order to see how the addition of features helps hyperbolic embedding we need to characterize the level of correlation between the three sets of labels. For this, we used the Normalized Mutual Information (NMI) to evaluate the agreement between the labels associated to each node in each class. Results are reported in Figure~\ref{fig:panel3} a,c,e and in Supplementary Figs.~7-8. There is always a significant amount of correlation between the labels. However, particular associations are more significant in specific real networks. For instance, all correlations are more important in Amazon Photo as compared to Wisconsin; in LastFM the association between classes and topology-based communities is higher than between classes and feature-based groups. Contrarily, Wisconsin exhibits the reverse trend, with classes being highly correlated with labels derived from features. 
		
		To evaluate the significance of the observed correlations, we calculate the z-score between two sets of labels. This measure compares the NMI of the original label sets with the NMI obtained when one set of labels is shuffled. A higher z-score indicates a more significant correlation between the two sets of labels (see Section~\ref{sec:random_labels}). In all the analyzed networks, the observed correlations are significant (z-score $>5$), except for the Texas and Cornell datasets, where the correlation between the classes and the topology-based communities is not statistically relevant.
		
		These correlations explain the performance of a node classification task defined on real network embeddings for the different labelling protocols (we omitted the evaluation of classifying feature-based labels in UMAP embeddings since those labels were crafted from that embedding). First, we split the nodes into training and test sets. The popular choice is 80\% of nodes into the training set and the rest 20\% into the test set~\cite{gholamy_kreinovich_kosheleva_2018}. Using the information of node labels in the training set, we predicted the labels of nodes in the test set using the KNeighborsClassifier from the scikit-learn library \cite{scikit-learn}. This neighbors-based classification method assigns the data category which has the most representatives within the $k$ nearest neighbors of the node. In our experiments, we set $k=5$ (we also checked $k=10$, which gave similar results) and used the angular distance between the nodes in the similarity subspace as the metric to define nearest neighbors. The results were averaged over $5$ train-test splits.
		
		Heatmaps showing the achieved accuracy in predicting the nodes' labels of Amazon Photo, LastFM and Wisconsin are shown in Fig.~\ref{fig:panel3} b,d,f. The results for the remaining networks are shown in Supplementary Fig.~11. When the classes are correlated with the topology-based communities, $FiD$-Mercator tends to improve the accuracy in assigning classes to nodes as compared to $D$-Mercator and UMAP (Fig.~\ref{fig:panel3} b,d,f, first column). Moreover, $FiD$-Mercator outperforms in correctly classifying topology-based labels (Fig.~\ref{fig:panel3} b,d, second column). On the contrary, when there is a lack of correlation between the two categories, $D$-Mercator performs better (Fig.~\ref{fig:panel3} f, second column). The $FiD$-Mercator embedding achieves higher accuracy when predicting labels constructed from features, as shown in Fig.~\ref{fig:panel3} b,d (third column). Furthermore, when classes correlate with feature-based labels, incorporating node features into the embedding method enhances accuracy, which aligns with expectations. This is illustrated in Fig.\ref{fig:panel3} b,d (first column). In all cases, UMAP achieves low accuracy predicting nodes' labels when the correlation between classes and feature-based groups is weak, and performs very poorly in classifying on topology-based communities. It is worth mentioning that for all embedding methods, the accuracy is higher than in randomized surrogates, in which the labels were shuffled, i.e., random assignment of the labels, while keeping the original number of communities. 
		
		All together, the experiments demonstrate a nuanced effect of incorporating feature data into network structure for node classification tasks based on geometric network embeddings. It can be detrimental when the correlation between features and topology is very low. However, when there is a high correlation between network connectivity and features, adding them significantly enhances the results.
		
\begin{figure}[t!]
			\centering
			\includegraphics[width=0.5\textwidth]{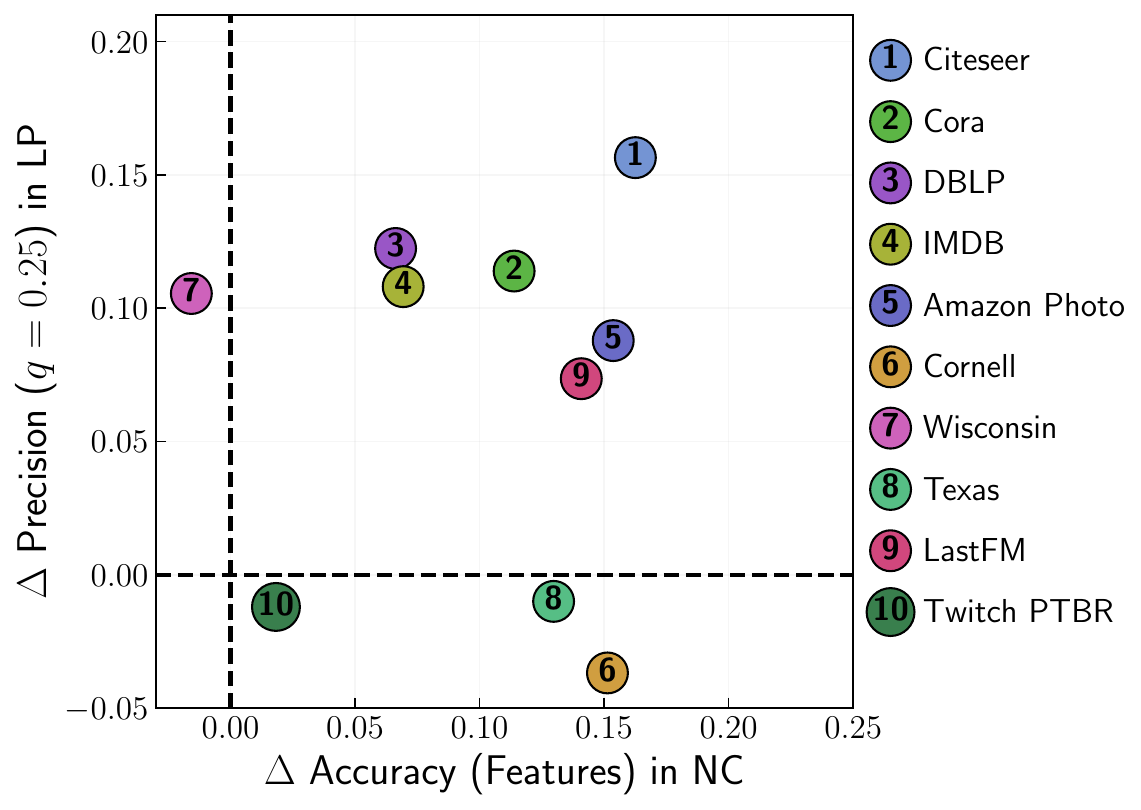}
			\caption{The difference of precision and accuracy between the $FiD$-Mercator and $D$-Mercator for real networks. For link prediction the precision was computed at 0.25 fraction of missing links. For node classification the accuracy was calculated for feature-based labels. Train-test size split: $80/20$. Results are averaged over $5$ different train-test splits.}
			\label{fig:panel4}
		\end{figure}

		\textbf{Versatility.} To clarify whether incorporating features into the embedding method provides an advantage in a variety of tasks, we show in Fig.~\ref{fig:panel4} the improvement in LP offered by $FiD$-Mercator over $D$-Mercator as a function of the improvement in NC using feature-based labels, for all datasets. In $60\%$ of the real networks the benefit is obvious in both tasks, while in three out of $10$ networks the results for LP, but not for NC, are slightly worse and in one network the result for NC, but not for LP, is slightly negative. Taken together, these results suggest that feature-aware embedding methods can improve a variety of downstream tasks beyond the ones explored here.
		
		\section{Discussion}
		Real-world complex networks are typically represented as graphs where connections are often enriched with attributes, such as the intensities of the interactions or their directionality. These attributes are incorporated in the network analysis process to obtain a better approximation to the actual behavior of the system. However, when node descriptive metadata different from connectivity is available, in the form of node features describing their qualities ( {\it e.g.} in social systems, age, gender, interests, background, and other social determinants), the common practice in network science is to not incorporating it into the analysis but to use the information as an external source of verification. For instance, in validating aggregation methods like community detection, determining node categories based on network structure. By contrast, advanced machine learning techniques in computer science, for instance many deep neural networks, incorporate information about both network structure and node features. In this case, the difficulty comes from the fact that the approaches are data-driven and the methods are black boxes in which the blending process becomes opaque and it is impossible to discern the precise role of every contributing factor.
		
		In this work, we surmount these drawbacks and merge network structure with node features in a model-based environment to obtain a more advantageous ultra-low dimensional reduction technique for real networks. This hybridization technique, that we named $FiD$-Mercator, produces $D$-dimensional geometric representations of real networks in hyperbolic space. The features determine a convenient initial condition to guide more efficiently the embedding of the network topology towards a better local optimum. The relative attention paid to the network structure versus the features can be controlled by early stoppage guided by the evolution of the log-likelihood function. As a result, the hybridization reveals a different type of spatial description by implementing a different inductive bias, that is, by considering a set of assumptions that involve both features and network structure simultaneously, and independently of whether those are or not correlated. Moreover, we found that adding node features to network structure does not disrupt the quality of the hyperbolic embedding in terms of its capacity at explaining local node properties, determining the degree distribution or the clustering spectrum. 
		
		However, its impact can vary depending on the correlation between the two and the relevance of each piece of information to a specific downstream task. In link prediction, the major improvement of $FiD$-Mercator is achieved in cases when the correlation is high but not perfect. Perfect correlation means that the information contained in the nodes' features would be redundant. Instead, high correlation below one implies that there is new information encoded in the features that can be used to infer structural properties of the network. When the correlation is very low, features and network connectivity are hardly related and the prediction of missing links cannot benefit from adding features. This suggests that there is an optimal correlation value for which better results are obtained. For node classification tasks, adding features can be detrimental when the correlation between features and topology is very low. In contrast, adding features significantly enhances the results when the correlation is high. 
		
		Nevertheless, there may be tasks where adding features is not optimal, independently of the level of correlation, which is difficult to know {\it a priori}. Other open questions for future research are how to smartly filter the set of features to adjust the correlation between features and network connectivity for optimal task performance, and the effects of incorporating non-binary features into the analysis, which could be easily mapped with UMAP. In any case, our results indicate that, overall, features are in general correlated with network connectivity and that the level of correlation is a good indicator of their relevance to improve many downstream tasks based on geometric network embeddings. This emphasizes the need for the development of new network embedding techniques that simultaneously optimize the joint probability of network structure and feature association in the near future.

		\section{Methods}
		\subsection{$\mathbb{S}^2$ model} 
		\label{sec:model}
		In the $\mathbb{S}^2$ model, a node $i$ is endowed with a hidden degree $\kappa_i$ and a position in the 2-similarity space, $\mathbf{v}_i =\{x_i, y_i, z_i\}$ with $||\mathbf{v}_i||=R$ where $R$ is the radius of the two-sphere. The connection probability between a node $i$ and a node $j$ has form:
		\begin{align}
			\label{eq:prob_conn}
			p_{ij} = \frac{1}{1 + {\chi_{ij}^\beta}}, \,\,\, \text{with} \,\,\, \chi_{ij} = \frac{R \Delta\theta_{ij}}{\sqrt{ \mu \kappa_i \kappa_j}}.
		\end{align}
		We fix the density of nodes in the two-sphere to one so that
		\begin{align}
			R = \sqrt{\frac{N}{4\pi}}
		\end{align}
		where $N$ is the number of nodes. The angular distance is defined as $\Delta\theta_{ij} = \arccos\left(\frac{\mathbf{v}_i \cdot \mathbf{v}_j}{R^2}\right)$. The parameter $\beta > 2$, named inverse temperature, controls the clustering in the network. Finally, the parameter $\mu$ controls the average degree of the network and is defined as
		\begin{align}
			\mu = \frac{\beta  \sin \frac{2\pi}{\beta}}{2\pi^2 \left<k\right>}.
			\label{eq:mu}
		\end{align}
		The hidden degrees can be generated randomly from an arbitrary distribution, or taken as a set of prescribed values. The model has the property that the expected value of the degree of a node with hidden variable $\kappa$ is $\bar{k}(\kappa) = \kappa$.

		\subsection{$FiD$-Mercator algorithm}
		\label{sec:mm}
		For a complete description of the inference of the hidden degrees and parameter $\beta$, the likelihood maximization and technique, and the final adjustment of hidden degrees please refer to~\cite{jankowski2023dmercator}.
		
		\textbf{Inferring the hidden degrees and parameter $\beta$.}
		The inference of the hidden degree and the inverse temperature $\beta$ is implemented as an iterative process. We start with the approximate guess for $\beta \in (2, 3)$ and initialize the hidden degrees as the observed degrees $\{k_i, i, \dots, N\}$ in the original network. 
		The estimation aims to adjust the hidden degrees such that the expected degree of each node in the model matches the observed degree in the original network. After the hidden degrees are computed, the theoretical mean of the local clustering coefficient of networks in $\mathbb{S}^2$ ensemble can be evaluated. If its value varies from the original network, $\bar{c}$, the value of $\beta$ is adjusted. Then the process is rerun using the current estimation of hidden degrees until a predetermined precision is reached.
		
		\textbf{Initial embedding of the nodes from UMAP.}
		The UMAP takes as the input the nodes' features matrix $N \times N_f$ with the distance metric. In our case, we use the Haversine metric, which is designed to calculate distances on a spherical surface. This process positions the nodes within the embedding space. For more details about the UMAP algorithm, see Section~\ref{sec:umap}.
		
		\textbf{Likelihood maximization.}
		The nodes' coordinates in the similarity space inferred using UMAP are fine-tuned by Maximum Likelihood Estimation (MLE) to optimize the probability that the observed network is generated by the $\mathbb{S}^2$ model.
		Nodes are visited sequentially by the network's onion decomposition~\cite{hebert2016multi}. For each node, we propose the candidate's positions near its neighbours. The most favorable proposed position, i.e., maximizing the local log-likelihood, is selected, and the process is repeated until the local log-likelihood function reaches a plateau.
		
		\textbf{Final adjustment of hidden degrees.}
		Lastly, we correct the hidden degree to compensate deviations from $\bar{k}(\kappa_i) = \kappa_i$, which might have been introduced in estimating the coordinates of nodes in the similarity spaces.
		
		\subsection{UMAP algorithm}
		\label{sec:umap}
		UMAP (Uniform Manifold Approximation and Projection) \cite{umap2018} is a dimension reduction technique. By default, it embeds data in Euclidean space. This is a convenient and widely used option for visualization and clustering \cite{Hao2021,Litvinukova2020}. However, there are no significant constraints that prevent the algorithm from working with non-Euclidean spaces.
		
		UMAP allows you to choose the dimension of the embeddings. We have verified that with $D=1$ performs poorly because we first need to map the nodes to a 2D plane and then place them on a circle. As this procedure mixes communities, we chose $D=2$ for UMAP dimension.
		
		UMAP maps the node feature vectors on top of the 2-sphere, which can be treated as a similarity space of the $\mathbb{S}^2$ model. It is worth noting that the UMAP algorithm is unsupervised, meaning it does not require any node labels.
		
		The UMAP algorithm, being stochastic, can produce different results in various runs. While the inferred positions of the nodes might slightly vary between these runs, our tests, involving running the UMAP algorithm twice, indicate that the obtained embeddings lead to consistent results in downstream tasks.
		
		\subsection{Correlation between nodes' features and network structure}
		\label{sec:correlation}
		Let us define a complementary graph of the network topology, $\mathcal{G}^\prime = (V, E^\prime)$ with $E^\prime \subseteq \{ (u, v) \,|\, u, v \in V \,\, {\textrm{and}} \,\, (u, v) \notin E \}$ where $V$ is the set of nodes in the original network and $E'$ is a set of edges non-existing in the original network. Computing cosine similarity between all non-existing links in $\mathcal{G}$ is computationally expensive due to its sparsity. As an alternative, we randomly selected $|E|$ edges from the complementary graph and averaged the results over five realizations of the selection process for further analysis.
		
		For each realization, we computed the average cosine similarity between connected nodes in graph $\mathcal{G}$ and between node pairs in the selected samples of $\mathcal{G}^\prime$. The relative difference between them quantifies the strength of the correlation between the nodes' features and the networks structure: 
		\begin{align}
			\mathrm{corr}(\mathcal{G}, F) = \frac{\overline{\mathrm{CS}}(\mathcal{G}) - \overline{\mathrm{CS}}(\mathcal{G}_E^\prime)}{ \overline{\mathrm{CS}}(\mathcal{G})}.	
		\end{align}
		The higher the obtained value the more correlated are features with the network topology.
		
		\subsection{Community concentration}
		\label{sec:concentration}
		Introduced in~\cite{jankowski2023dmercator}, the geometric concentration of a community $l$ around a node $i$ has a form
		\begin{align}
			\rho_{i, l} = \frac{n_{i, l}}{n_{i, g}}\frac{N}{N_L},
		\end{align}
		where $n_{i, l}$ is the number of nodes in community $l$ out of $n_{i, g}$ considered nodes, and $N_L$ is the total number of nodes in community $l$. We define $n_{i, g}$ as top geometrically closest neighbors. The single scalar $\rho$ for a given network embedding is averaged over all nodes in the same community. Finally, we calculated the community concentration as $c_C = \rho(n_{i, l} = N/10)$, i.e., the geometric concentration at $10\%$ of top geometrically closest nodes.
		
		\subsection{z-score}
		\label{sec:random_labels}
		To measure the significance of the correlation between two sets of labels, we used z-score. Let us define the quantity
		\begin{align}
			\mathrm{z}{\text -}\mathrm{score} = \frac{\mathrm{NMI(X, Y)} - \overline{\mathrm{NMI(X, Y)}}_{\text{rand}}}{\sigma(\mathrm{NMI(X, Y)}_{\text{rand}})},
		\end{align}
		where $X$ and $Y$ are sets of labels. We compute the random case of by comparing set $X$ with a shuffled version of set $Y$, and this process is repeated $100$ times. The higher z-score the more significant is the correlation between two sets of labels.
		
		\subsection{Datasets}
		\label{sec:datasets}
		In this work we analyzed $10$ networks from different domains. Their properties are summarized in Table~\ref{tab:1}. For each dataset, the nodes' feature vector is a high-dimensional vector containing $0$s or $1$s, i.e., the presence or absence of a given feature.
		
		\begin{table*}[!htbp]
			\centering
			\begin{adjustbox}{width=0.85\textwidth} 
					\begin{tabular}{lccccccccccc}
							\hline
							\hline
							Network      & $N$    & $\left<k\right>$     & $\bar{c}$    & $N_F$ & $\left<k_F\right>$ & $N_L^{\mathrm{Class}}$ & $N_L^{\mathrm{Topology}}$ & $N_L^{\mathrm{Features}}$ & $\beta$ & $\mu$ & $\mathrm{corr}(\mathcal{G}, F)$\\ \hline
							Citeseer     & 2110 & 3.48  & 0.23 & 3703  & 32.06  & 6  & 37 & 4 & 2.98 & 0.0375 & 0.763 \\ 
							Cora         & 2485 & 4.08  & 0.28 & 1433  & 18.30  & 7  & 27 & 5 & 3.23 & 0.0373 & 0.650 \\
							DBLP         & 2728 & 31.44 & 0.67 & 2000  & 42.14  & 4  & 18 & 4 & 7.21 & 0.0089 & 0.506 \\ 
							IMDB         & 3228 & 19.46 & 0.55 & 2000  & 76.95  & 3  & 27 & 5 & 5.12 & 0.0125 & 0.172 \\ 
							Amazon Photo & 7487 & 31.8  & 0.42 & 745   & 260.47 & 8  & 14 & 7 & 3.41 & 0.0052 & 0.270 \\ 
							Cornell      & 184  & 3.03  & 0.29 & 1703  & 94.20  & 5  & 14 & 4 & 2.40 & 0.0200 & 0.169 \\ 
							Wisconsin    & 251  & 3.59  & 0.28 & 1703  & 95.84  & 5  & 12 & 5 & 2.49 & 0.0204 & 0.201 \\ 
							Texas        & 183  & 3.05  & 0.32 & 1703  & 83.42  & 5  & 12 & 4 & 2.24 & 0.0123 & 0.116 \\
							LastFM       & 7624 & 7.29  & 0.29 & 7842  & 395.38 & 18 & 29 & 5 & 2.96 & 0.0175 & 0.465 \\
							Twitch PTBR  & 1912 & 32.74 & 0.34 & 3169  & 19.88  & 2  & 8  & 5 & 2.39 & 0.0018 & -0.053 \\
							\hline
							\hline
						\end{tabular}
				\end{adjustbox}
			\caption{Properties of real networks. The $N_F$ represents the number of features, $\left<k_F\right>$ the average number of features per node and $N_L$ the number of labels. We used the giant connected component for all networks.}
			\label{tab:1}
			\end{table*}
		
		\begin{itemize}
			\item Citeseer \cite{giles1998citeseer}: The citation network of Machine Learning papers where each publication is described by a $0$ or $1$ valued word vector indicating the absence or the presence of the corresponding word from the dictionary. The dictionary consists of $1433$ unique words. The publications are classified into six classes.
			\item Cora \cite{McCallum2000}: Similar to Citeseer, however the publications are split into seven classes: Case Based, Genetic Algorithms, Neural Networks, Probabilistic Methods, Reinforcement Learning, Rule Learning, Theory.
			\item DBLP \cite{wang2019heterogeneous}: The citation network of Machine Learning papers. Nodes are classified into four classes: Data Mining, Artificial Intelligence, Computer Vision and Natural Language Processing.
			\item IMDB \cite{wang2019heterogeneous}: The Movie-Actor-Movie relation dataset. Movies are categorized into three classes (Action, Comedy, Drama). 
			\item Amazon Photo \cite{shchur2018pitfalls}: Nodes represent goods and edges represent that two goods are frequently bought together. The node features are bag-of-words encoded product reviews, and class labels are given by the product category.
			\item Cornell, Wisconsin, Texas \cite{Craven1998}: Web graphs crawled from three Computer Science departments in $1998$, with each page manually classified into one of seven categories: course, department, faculty, project, staff, student, or other.	
			\item LastFM \cite{feather}: A social network collected from the public API in March $2020$. Nodes are LastFM users from Asian countries and edges are mutual follower relationships between them. The vertex features are extracted based on the artists liked by the users. This node label was derived from the country field for each user.
			\item Twitch PTBR \cite{musae}: User-user networks of gamers who stream in Portuguese language. Nodes are the users themselves and the links are mutual friendships between them. Vertex features are extracted based on the games played and liked, location and streaming habits. The node labels represent whether a streamer uses explicit language.
		\end{itemize}

		\section*{Data availability}
		The network datasets used in this study are available from the sources referenced in the manuscript and the supplementary materials.
		
		\section*{Code availability}
		The open-source code of $FiD$-Mercator is available on GitHub at \url{https://github.com/networkgeometry/FiD-Mercator}.
		
		\section*{Acknowledgments}
		We acknowledge support from: Grant TED2021-129791B-I00 funded by MCIN/AEI/10.13039/501100011033 and the ``European Union NextGenerationEU/PRTR''; Grant PID2019-106290GB-C22 funded by MCIN/AEI/10.13039/501100011033; Generalitat de Catalunya grant number 2021SGR00856. R.~J. acknowledge support from the fellowship FI-SDUR funded by Generalitat de Catalunya. P.~H. acknowledges support from the NSF AccelNet-MultiNet program under Grant No. 1927425 at Northeastern University. M.~B. acknowledges the ICREA Academia award, funded by the Generalitat de Catalunya.
		
		\section*{Author contributions}
		R.J. implemented the methods. M.A.S. supervised the work and wrote the draft of the paper. All authors contributed to the research design, result analysis, and paper writing.
		
	

	\includepdf[pages={{},{},1,{},2,{},3,{},4,{},5,{},6,{},7,{},8,{},9,{},10,{},11,{},12,{},13,{},14,{},15,{},16,{},17,{},18,{},19,{},20,{},21,{},22,{},23,{},24}]{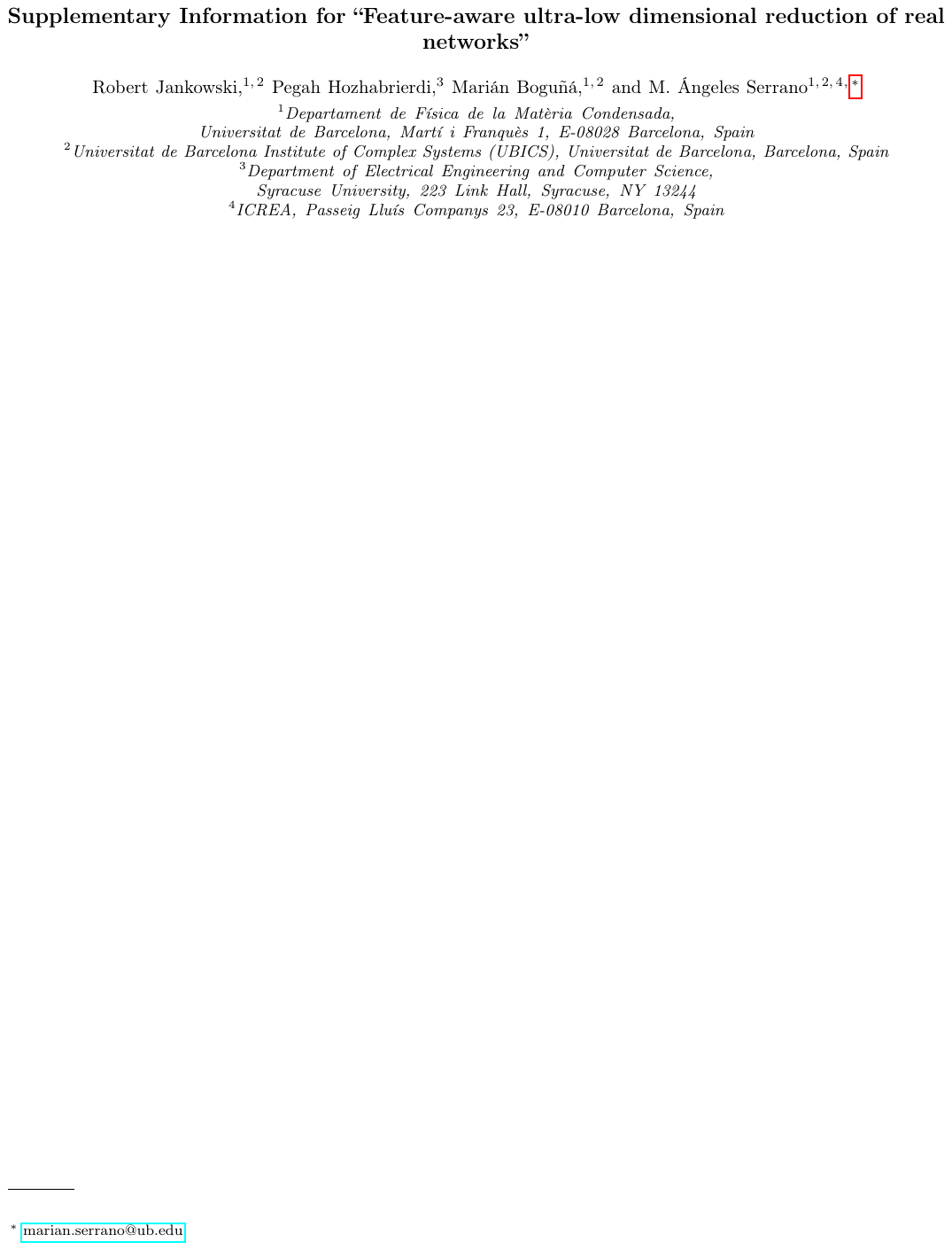}
	
\end{document}